\documentclass[usenatbib]{mn2e}
\usepackage{natbib}
\usepackage{graphicx}
\usepackage{epsfig}
\usepackage{amsmath}
\title{The Impact of Stellar Model Spectra in Disk Detection}
\author[J. A. Sinclair, Ch. Helling, J. S. Greaves]
       {J. A. Sinclair, Ch. Helling, J. S. Greaves \\
        SUPA, School of Physics and Astronomy, University of St. Andrews, North Haugh, St. Andrews, Fife, United Kingdom, KY16 9SS}
\date{Accepted 2010 July ?.
      Received 2010 July ?;
      in original form 2010 July}

\pagerange{\pageref{firstpage}--\pageref{lastpage}}
\pubyear{2010}

\begin{document}

\maketitle

\label{firstpage}

\begin{abstract}

\noindent We present a study of the impact of different model groups
in the detection of circumstellar debris disks.  Almost all previous
studies in this field have used {\sc kurucz} ({\sc atlas9}) model
spectra to predict the stellar contribution to the flux at the
wavelength of observation thus determining the existence of a disk
excess.  Only recently have other model groups or \emph{families} like
{\sc marcs} and {\sc {\sc NextGen}} ({\sc phoenix}) become available
to the same extent as {\sc atlas9}.  This study aims to determine
whether the predicted stellar flux of a disk target can change with
the choice of model family - can a disk excess be present in the use
of one model family whilst being absent from another?  A simple
comparison of {\sc kurucz} model spectra with {\sc marcs} and {\sc
NextGen} model spectra of identical stellar parameters was conducted
and differences were present at near-infrared wavelengths.  Model
spectra often do not extend in wavelength to that of observation and
therefore extrapolation of the spectrum is required.  In extrapolation
of model spectra to the Spitzer MIPS passbands, prediction of the
stellar contribution differed by $5$ \% at $70$ $\mu$m for F, G and
early K spectral types with differences increasing to $15$ \% for
early M dwarfs.  Analysis of the Spitzer MIPS $24$ $\mu$m observations
of 37 F, G and K solar-like stars in the Pleiades cluster was
conducted.  In using {\sc kurucz} model spectra, 7 disk excesses were
detected while only 3 and 4 excesses were detected in using {\sc
marcs} and {\sc NextGen} ({\sc phoenix}) model spectra respectively.

\end{abstract} 
\section{Introduction}

\noindent The study of Protoplanetary and Debris Disks is a
significant area of research today.  It is uncertain whether the
processes which formed our own solar system are typical of other
planetary systems.  Both protoplanetary and debris disks
describe planetary systems in the midst of formation/evolution and
therefore are crucial to our understanding of planetary systems as a
whole.  
In steady state, the mutual collisions of planetesimals replenish dust
removed from the disk by radiation pressure and Poynting-Robertson
drag.  Planetesimals and regenerated dust around older stars describe
debris disks in contrast to the first-generation dust around stars
younger than 10 Myr old.

\noindent Detection of a disk may be gained if infrared emission from
an object is in excess of expectation from the stellar flux.  This is
determined numerically by fitting a model spectrum and extrapolating
out to the wavelength of observation - see the Appendix of
\cite{bryden06} for details of this procedure.  The excess ratio
($F_{\text{obs}}/F_{\text{pred}}$) is computed and detection of an
infrared excess is concluded when this value exceeds the 3$\sigma$
limit of the entire sample (\cite{aumann_discovery_1984}, \cite{beichman_planets_2005}).  In almost all past and
present studies of debris disks, {\sc kurucz} ({\sc ATLAS9}) models
(\cite{castelli2004}) are used for this procedure.  Other model
groups like {\sc marcs} (\cite{gustaf08}) and {\sc NextGen} ({\sc
phoenix}) (\cite{haus1999}) remain largely unused and have only recently become readily available.

\noindent Standard 1D model spectra are the result of a solution of
the equation of radiative transfer, the hydrostatic equation,
mixing-length theory and the equations describing the gas phase in
chemical equilibrium.  Ideally, every possible atomic and molecular
transition should be included however this is computationally
impractical and simplifications are necessary.  {\sc marcs} and {\sc
NextGen} model spectra are produced using Opacity Sampling (\cite{os})
where opacities are distributed statistically in wavelength giving
rise to high resolution model spectra.  {\sc kurucz} model spectra
however are produced using Opacity Distribution Functions (ODFs) where
absorption cross-sections are binned producing a smoother, lower
resolution model spectrum (\cite{odf1}, \cite{odf2}).  Discrepancies
in line features are expected between model groups due to these
different opacity treatments though such differences are not expected
at the longer wavelengths at which disks are observed.   The
different model groups include opacities drawn from different
literature.  This is likely to introduce sizeable differences in the
spectral appearance of the model groups; for a discussion, see
\cite{jorgensen03}.  Model spectra with identical stellar parameters
but of different model families are compared in Section 2 of this
paper to determine whether such differences exist.  {The model
spectra often do not extend to longer wavelengths at which
observations of disks are made therefore, fitting of a Rayleigh-Jeans
tail and extrapolation to the required wavelength is necessary
(Section 3).  {\sc NextGen} models do extend beyond $100$ $\mu$m
however, the sparse wavelength distribution at these wavelengths make
it necessary to fit a Rayleigh-Jeans tail for interpolation onto some
filter bandpass.  Section 4 applies the use of all three model
families to 37 solar-like stars previously analysed for disk excesses
in \cite{sierchio10}.

\begin{figure}
  \centering
  \includegraphics[scale=0.38,angle=90]{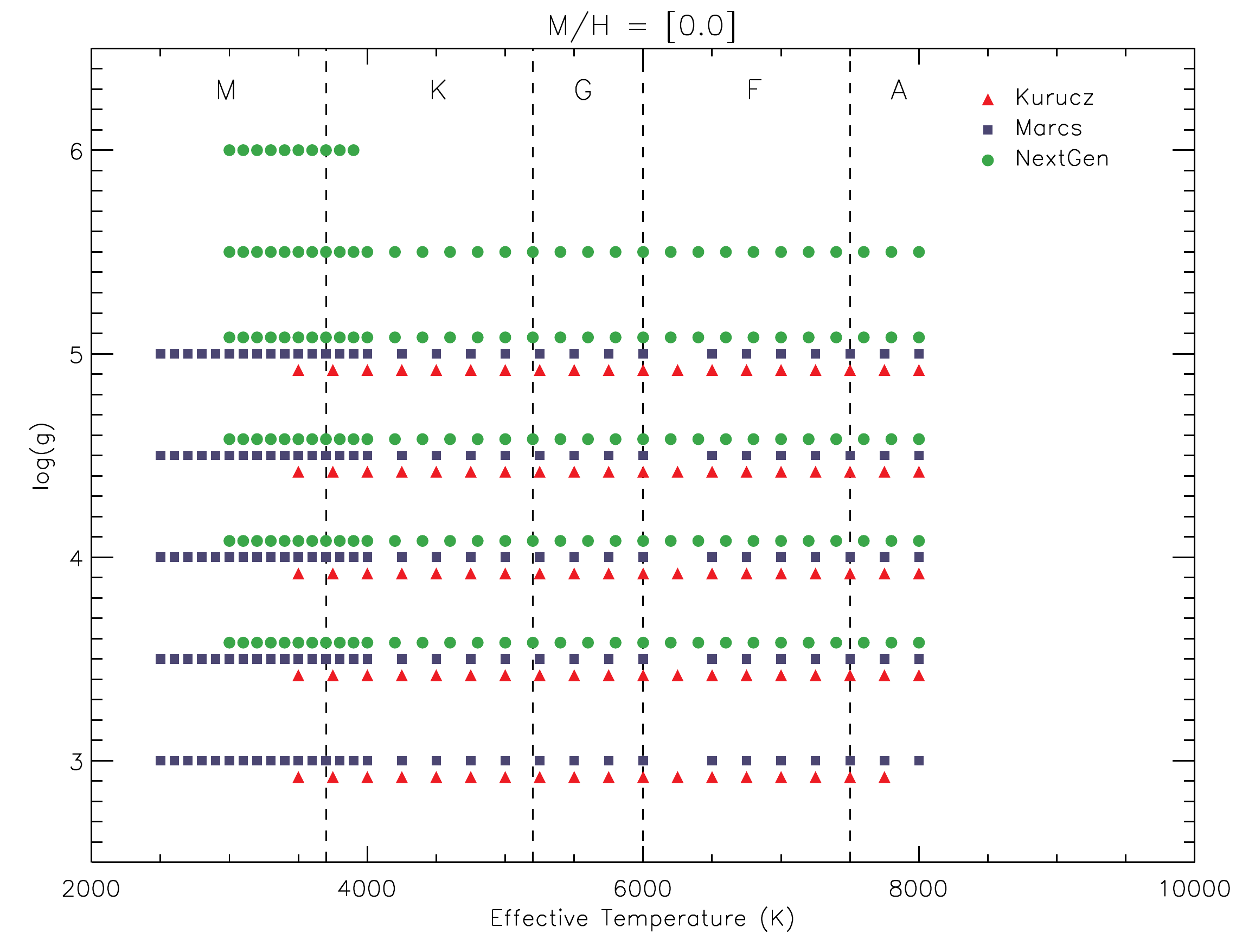}
    \caption{The available effective temperature and surface gravity
    combinations for {\sc kurucz} (red triangles), {\sc marcs} (blue
    squares) and {\sc NextGen} (green circles) spectra of solar
    metallicity.  The ranges of effective temperature which approximately represent
    the A, F, G, K and M spectral types are also displayed.}
    \label{fig:tefflogg}
\end{figure}

\vspace{-0.75cm}
\section{Comparison of Model Spectra}

\noindent Spectra of all three model families were compiled.  {\sc
kurucz}\footnote[1]{http://kurucz.harvard.edu/grids.html} and {\sc
NextGen}\footnote[2]{ftp://phoenix.hs.uni-hamburg.de/NextGen/Spectra/}
model spectra with inclusion of a $2.0$ $km$ $s^{-1}$ microturbulence,
a convective mixing length, $l/H$ = 1.25 and no parameterization of
convective overshoot were chosen (\cite{castelli2004},
\cite{haus1999}).  {\sc
marcs}\footnote[3]{http://marcs.astro.uu.se/search.php} model spectra
with identical microturbulence but (the only available at the time)
convective mixing, $l/H$ = 1.5 were chosen (\cite{gustaf08}).  {\sc
kurucz} and {\sc NextGen} model groups provide accurate synthetic SEDs
of solar-like stars though {\sc kurucz} better reproduces early-type
stars while {\sc NextGen} is more accurate in the M regime
(\cite{atlasnextgen}).  The Marcs model group have also been shown to
adequately match the stellar SEDs given their use in the photometry
calibration of the Spitzer Space Telescope (\cite{marcsspitzer}).
Fig. \ref{fig:tefflogg} displays the effective temperature and surface
gravity combinations of all three model families.  As shown, model
spectra in all three families adequately span the F, G and K spectral
types which dominate the stellar sample of disk surveys.  Model
spectra of solar metallicity are considered the most appropriate in
the context of disk surveys using stellar samples of solar-like, solar
neighbourhood stars which typically exhibit metallicities of [Fe/H] =
0.25 (\cite{valenti_spectroscopic_2005}).

\noindent A comparison of the model spectra across the model
families was conducted.  A {\sc kurucz} model spectra of some
effective temperature, surface gravity (and solar metallicity) was
compared with a {\sc marcs} and {\sc NextGen} model spectra with
identical stellar parameters.  The wavelength range
and resolution differed between model families therefore interpolation
(Eq. \ref{eq:li}) was necessary.  In comparing {\sc kurucz} and {\sc
marcs} spectra, the former were interpolated onto the wavelength
distribution of the latter.  { For comparison with {\sc NextGen} models,
a few short wavelength points were omitted to allow a common wavelength range and the {\sc kurucz} spectra
interpolated onto the {\sc NextGen} wavelength distribution}.  A ratio
of the flux (\ref{eq:ratio}) was computed.
\begin{equation}
\label{eq:li}
{F_i}' = F_{i-1} + (\lambda_j - \lambda_{i-1})\frac{(F_{i}-F_{i-1})}{(\lambda_{i} - \lambda_{i-1})}
\end{equation}
\vspace{-0.4cm}
\begin{equation}
\label{eq:ratio}
F_{\text{ratio}} (\lambda) = \frac{F_j}{{F_i}'}
\end{equation}
\noindent where $\lambda_i$, $F_i$ represents the {\sc kurucz} model spectrum while $\lambda_j$, $F_j$ represents the {\sc marcs} or {\sc NextGen} model spectrum where $\lambda_j$ is a wavelength intermediate to $\lambda_{i-1}$ and $\lambda_{i}$.   $F_{\text{ratio}} (\lambda)$ was calculated for each set of synthetic spectra and subsequently averaged for all spectra of the F, G and K spectral types.

\begin{figure*}
  \centering
{\label{fig:kurmar}
\includegraphics[width=0.35\textwidth,angle=90]{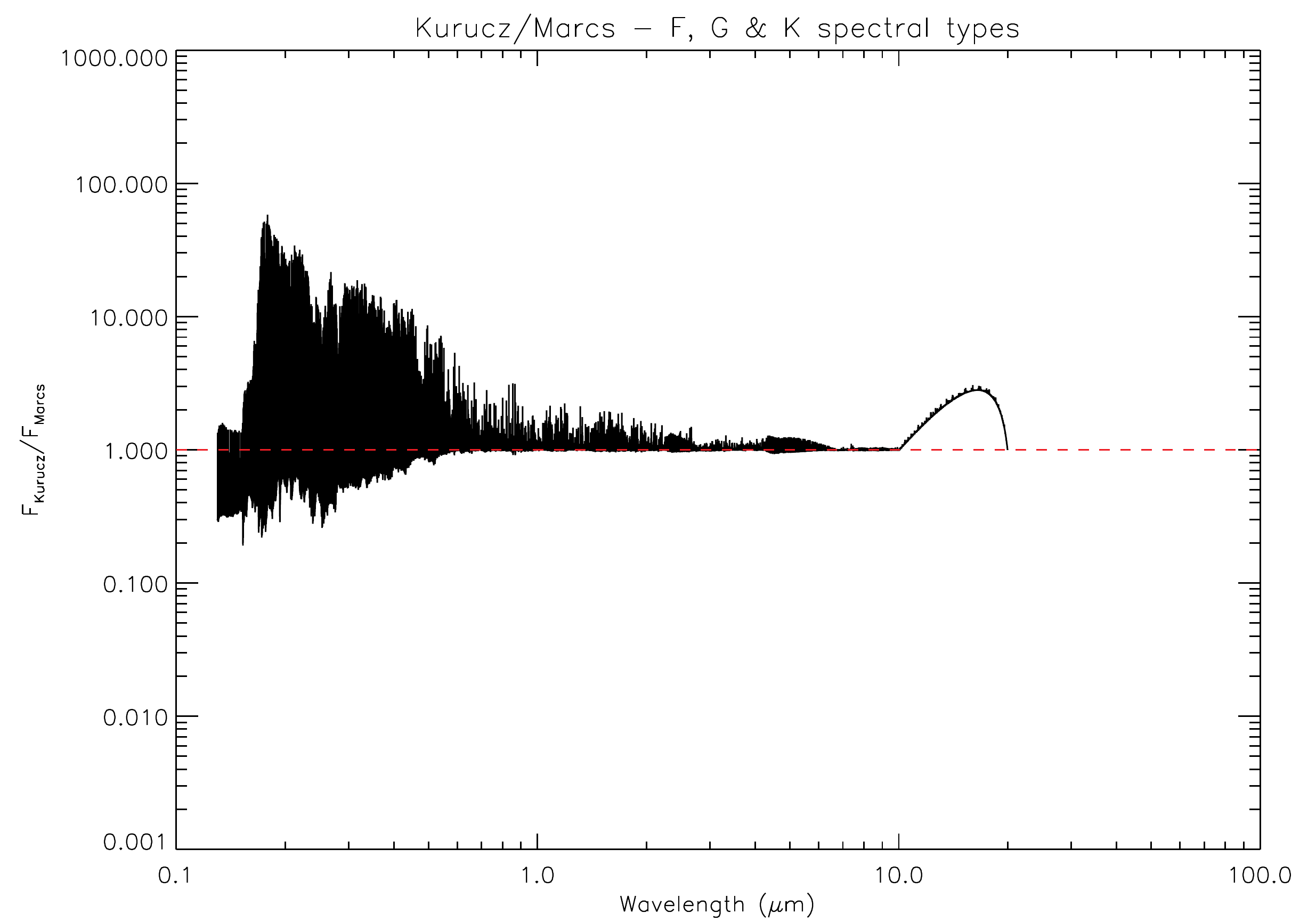}}                
{\label{fig:kurng}
\includegraphics[width=0.35\textwidth,angle=90]{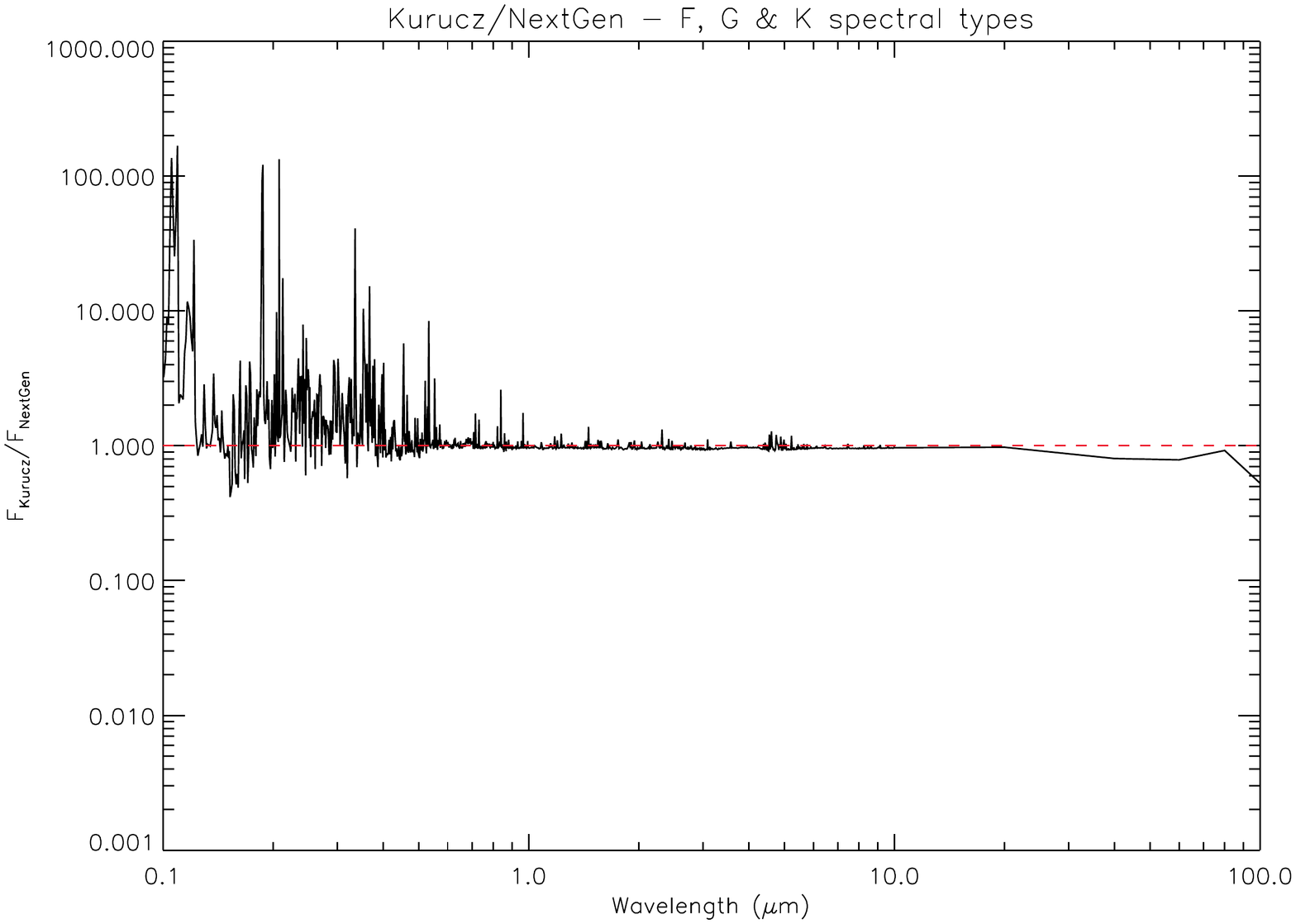}}
  \caption{The ratio of the {\sc kurucz} spectral flux with (a) {\sc marcs} flux and (b) {\sc NextGen} flux against wavelength ($\mu$m) averaged for all log(g) = 4.5, solar metallicity model spectra of F, G and K spectral types.  The scales are logarithmic and the dashed red line represents perfect agreement.}
\end{figure*}

\noindent Spectral fluxes of {\sc kurucz} and {\sc marcs} spectra can
differ by greater than an order of magnitude (Fig. \ref{fig:kurmar}).
In particular, the poorest agreement occurs at approximately $0.2$
$\mu$m.  The trend of the flux ratio is dominated by
delta-function-like features which likely are produced as a result of
line features present in one model group but absent from the other.
This is likely capturing the different opacity treatments.  A higher level of agreement occurs between {\sc kurucz} and
{\sc NextGen} model spectra (Fig. \ref{fig:kurng}).  Line features remain present, again
capturing the different opacity treatments.  The trends of both comparison tend to unity with increasing
wavelength showing better agreement of the synthetic spectra at longer
wavelengths.  The best agreement is apparent for the {\sc kurucz} and
{\sc NextGen} models while there exist some fluctuation about unity
for the {\sc kurucz}/{\sc marcs} comparison. {The
bump-like feature at approximately $10$ $\mu$m in the {\sc Kurucz}/{\sc Marcs} comparison possibly arises from differences in H$^-$ opacity, the dominant source at these wavelengths.  The deviation from unity in the Kurucz/NextGen comparison arises from an unphysical anomaly present in the {\sc Kurucz} T = 8000 K, log(g) = 4.5, [M/H] = 0.0 model spectrum.  This feature was masked in all subsequent analysis}.  These differences in the model spectra has obvious implications in
stellar flux prediction and finding the excess ratio from which
detection of a disk is decided.
\vspace{-0.5cm}
\section{Extrapolation to longer wavelengths}

\noindent The algebraic form of the Rayleigh-Jeans' flux ($F_{RJ}
(\lambda,T) = 2kTc/{{\lambda}^4})$ did not provide a good fit of the
long wavelength regions of the synthetic spectra.  $F_{RJ}
(\lambda,T_{\text{eff}})$ (where $T_{\text{eff}}$ is the effective
temperature of the model spectrum) was offset from the synthetic
spectrum by up to $40$\% at $5$ $\mu$m.  A simple scaling of $F_{RJ}$ to the
synthetic spectrum did not offer a solution as the fall-off of the
synthetic flux with wavelength deviated from $\lambda^{-4}$.  Such an
offset and slope difference remained when the full blackbody flux was
instead fit.  A straight
line fit of the spectrum in logarithmic space was instead conducted.  Eq. \ref{eq:logf} was fit (allowing freedom of
$a$ and $b$ which minimised $\chi^2$) and transformed back into normal
space, giving a function of the form shown in Eq. \ref{eq:flam}.
\begin{equation}
\log(F_{\lambda}) = a \, \log(\lambda) + b
\label{eq:logf}
\end{equation}
\begin{equation}
F_{\lambda}=\frac{10^b}{\lambda^a}
\label{eq:flam}
\end{equation}

\noindent where $a$ describes the wavelength power index
(approximately 4) while $b$ serves as a scale for the pattern.  Significant differences in the parameters which minimised $\chi^2$ were apparent - for example, $a \approx 3.9$, $b \approx 6.66$ provides a good fit for a {\sc kurucz} T = 3500 K, log(g) = 4.5 and [M/H] = 0.0 model spectrum while $a \approx 4.03$, $b \approx 6.94$ provides a good fit for a {\sc NextGen} model spectrum with the same parameters.  For each model spectrum, such a fit was performed, the
spectrum extrapolated out to $24$ $\mu$m and $70$ $\mu$m and the flux
integrated over the corresponding Spitzer MIPS passbands which are
sensitive to dust emission at 1 AU and outside 10 AU from a solar-like
star respectively (\cite{rieke_multiband_2004}).  For each set of {\sc kurucz} and {\sc marcs} or {\sc
NextGen} model spectra identical in stellar parameters, the broadband
fluxes of these passbands were compared.  Fig. \ref{fig:compspitz}
displays a ratio of these fluxes for each set of synthetic spectra.  Kurucz model spectra do not extend to the cooler effective temperatures of M dwarfs and therefore Marcs and NextGen model spectra are also compared in this way to determine the possible differences in flux prediction for such objects.  

\begin{figure*}
  \centering
  \includegraphics[scale=0.75,angle=90]{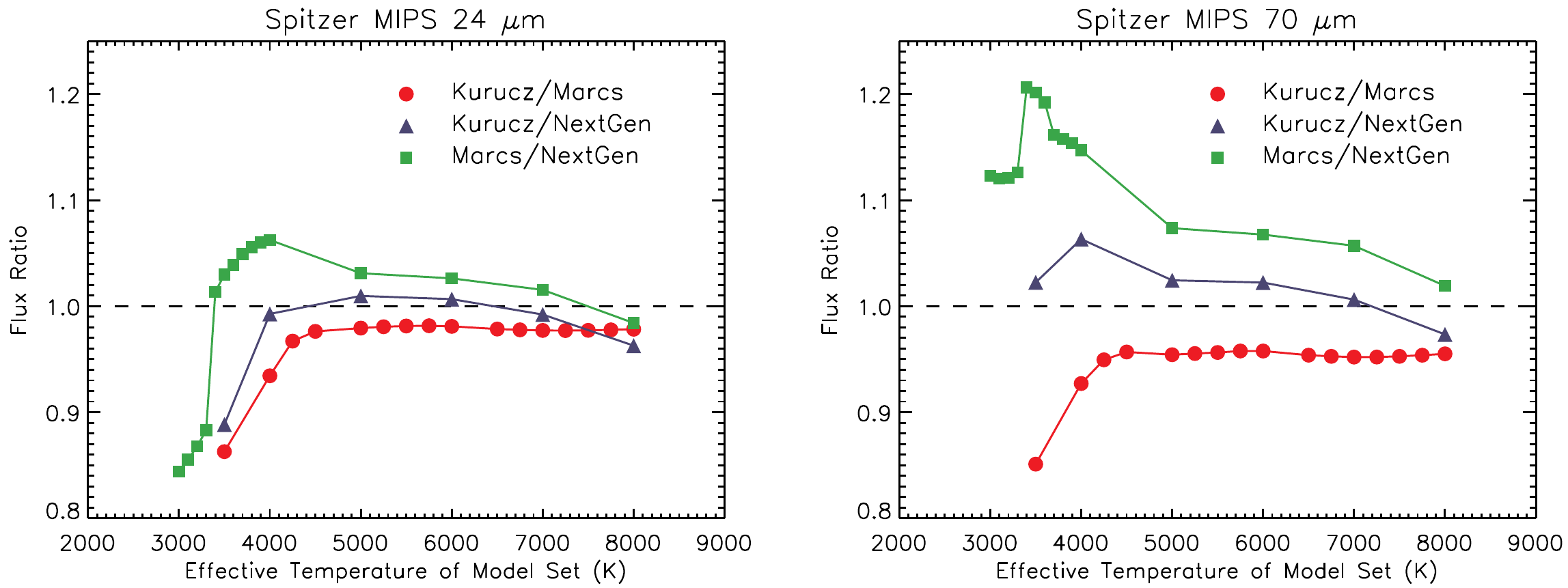}
    \caption{A comparison of the Spitzer MIPS predicted fluxes against
    the effective temperature of the log(g) = 4.5 {\sc kurucz} and
    {\sc marcs}/{\sc NextGen} model spectra compared.  The {\sc kurucz}/{\sc
    marcs} comparison is shown as blue triangles, the {\sc
    kurucz}/{\sc NextGen} comparison are represented by red circles and the Marcs/NextGen comparison shown as the green squares.
    There are comparably fewer points for the {\sc kurucz} and {\sc
    NextGen} comparison as fewer spectra had matching stellar
    parameters.  The dashed horizontal lines mark agreement of the fluxes.}
    \label{fig:compspitz}
\end{figure*} 

\noindent At $24$ $\mu$m, the extrapolated fluxes of {\sc kurucz} and
{\sc marcs} model spectra exhibit good agreement (of approximately 2
\%) for F, G and early K spectral types.  The trend deviates
significantly from unity for cooler late K and early M spectral types
with disagreement as high as $15$ \% in {\sc kurucz} and {\sc marcs}
spectra of 3500 K.  At $70$ $\mu$m, the flux ratio of {\sc kurucz} and
{\sc marcs} model spectra is identical to that of $24$ $\mu$m though
it has a larger deviation from unity (approximately $5$\%).  This
suggests that the long wavelength regions of {\sc kurucz} and {\sc
marcs} synthetic spectra differ in slope and therefore their log-log
straight line fits (Eq. \ref{eq:logf}) diverge with wavelength.  In
comparison, the {\sc kurucz}/{\sc NextGen} trend exhibits a higher
level of agreement which is consistent with Fig. \ref{fig:kurng}.  At
$24$ $\mu$m, the flux ratio almost perfectly centres on unity with
large disagreement only apparent in the comparison of early M dwarfs.
At $70$ $\mu$m, a small deviation from unity is apparent though
agreement is still true within $3$ \% for F, G and K spectral types.
Similarly, {\sc Marcs} and {\sc NextGen} model spectra of F, G and K
spectral types exhibit good agreement (of approximately 4 \%) at $24$
$\mu$m, deviating to approximately $8$ \% at $70$ $\mu$m.  Again, M
dwarfs exhibit the worst agreement with differences as large as $20$
\%.  The poor agreement of M dwarf model spectra likely arises from
the challenges in determining the opacity-relevant effects of
molecular species present in such atmospheres (\cite{gustafmols},
\cite{jorgensen03}, see also section 2.2 of \cite{hellinglucas}).
Observational studies have noted discrepancies between the observed
and synthetic colours of such objects
(\cite{gautier_far-infrared_2007}).
\vspace{-1.0cm}
\section{Analysis of 37 solar-like Pleiades stars}

\noindent The possible impact in the use of different model groups in
flux prediction of disk observations has so far only been inferred
from comparison of the model spectra.  It is therefore ideal to
physically apply using all three model spectra to a set of
observations.

\noindent A recent publication (\cite{sierchio10}) analysed the
Spitzer MIPS $24$ $\mu$m observations of approximately 70 stars in the
Pleiades cluster.  Using {\sc kurucz} model spectra for prediction
of the stellar flux at $24$ $\mu$m, they concluded that 23 of such
stars harboured disks.  Using a smaller subset of their target sample
(due to limitations in the available photometry) this study was
reproduced using {\sc kurucz}, {\sc marcs} and {\sc NextGen} model spectra.

\noindent Johnson B, V, R, I and 2MASS J, H, Ks photometry of each
object was compiled where possible.  Synthetic photometry of all
spectra of all three model groups was searched and $\chi^2$ calculated
in each instance.  The observed and synthetic photometry differed
significantly by the target distance modulus.  Some targets in the
Pleiades sample had known accurate (Hipparcos) distances however, for
consistency and to avoid further reducing the stellar sample of the
stars to those only with parallax measurements, an alternative method
was decided.  A mean magnitude in the above bands was calculated for
both the synthetic and observed photometric magnitudes.  The
difference in these mean values was considered to give a value of the
distance modulus.  U band photometry was highly offset from longer
wavelength photometry, as noted in the Appendix of \cite{bryden06}, and was
therefore omitted in this procedure.  Chromospheric activity was
considered to be the most likely explanation for these offsets in the
U band (T. Lloyd-Evans, A. C. Cameron, private communication, 2010).
Having found the synthetic spectrum in all three model families which
provided the best photometry fit, the `distance modulus' was added to
the synthetic Spitzer MIPS $24$ $\mu$m magnitude to determine the
predicted flux.  

\noindent For each object, the predicted fluxes using all three model
families were compared.  Fig. \ref{fig:bland} displays a plot of the
predicted fluxes of the 37 objects using {\sc kurucz} model spectra against that of {\sc
marcs} and {\sc NextGen} models.  As shown, the line of equality
provides an adequate best-fit of each data set.  Although many points
lie very close to equality, there are also many outliers and
therefore, flux predictions for some objects using different model families can be
significantly different.

\begin{figure}
  \centering  \includegraphics[scale=0.6]{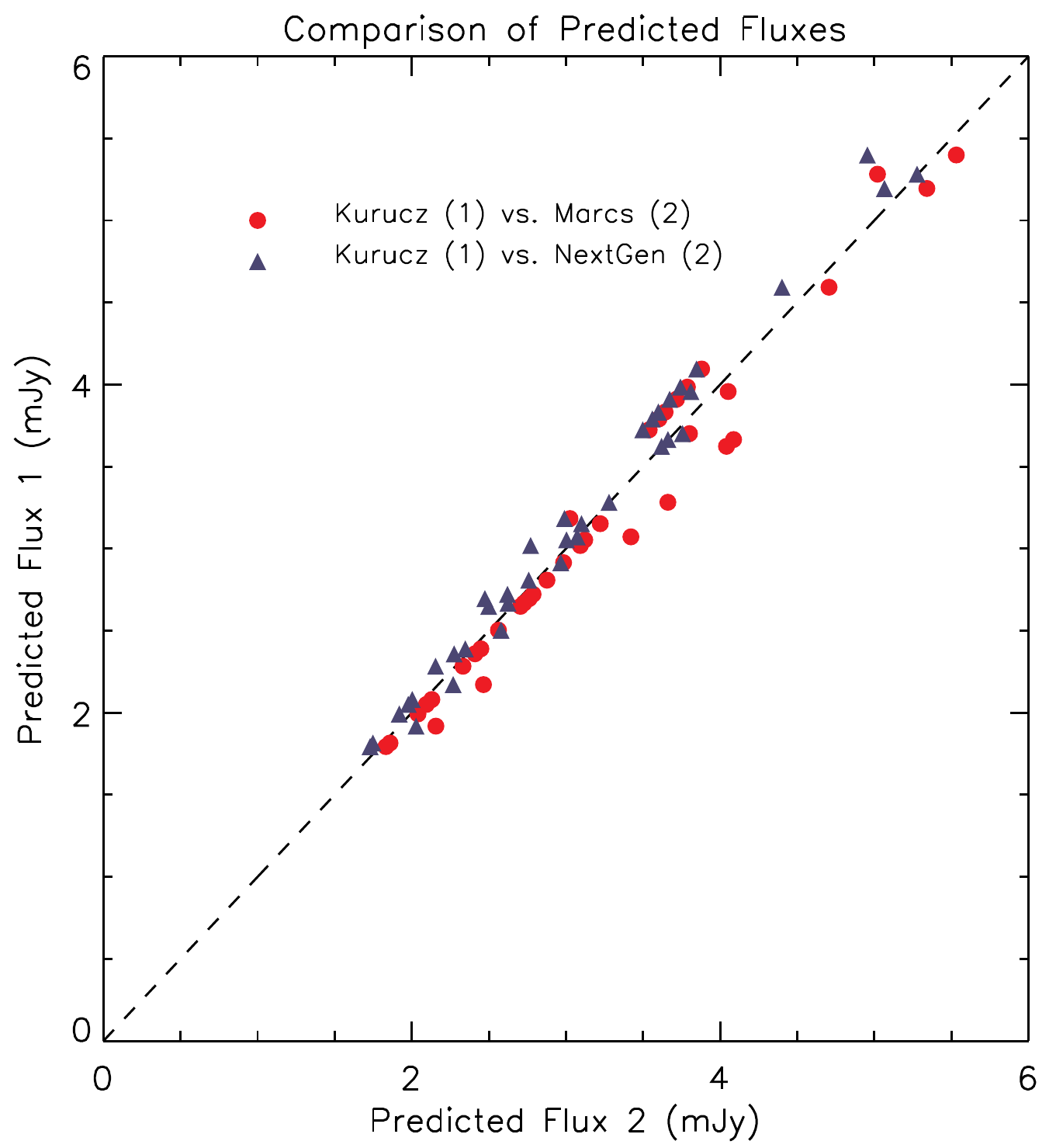}
    \caption{A comparison of the predicted fluxes of the 37 targets
    (\protect\cite{sierchio10}) at $24$ $\mu$m (in mJy) in using {\sc kurucz},
    {\sc marcs} and {\sc NextGen} model spectra.  Red circles show
    the comparison for {\sc kurucz} and {\sc marcs} models while blue
    triangles show that of {\sc kurucz} and {\sc NextGen} models.  The
    dashed line represents equality and is not a fit of either data.
    set.}
    \label{fig:bland}
\end{figure}

\begin{figure}
  \centering \includegraphics[scale=0.8]{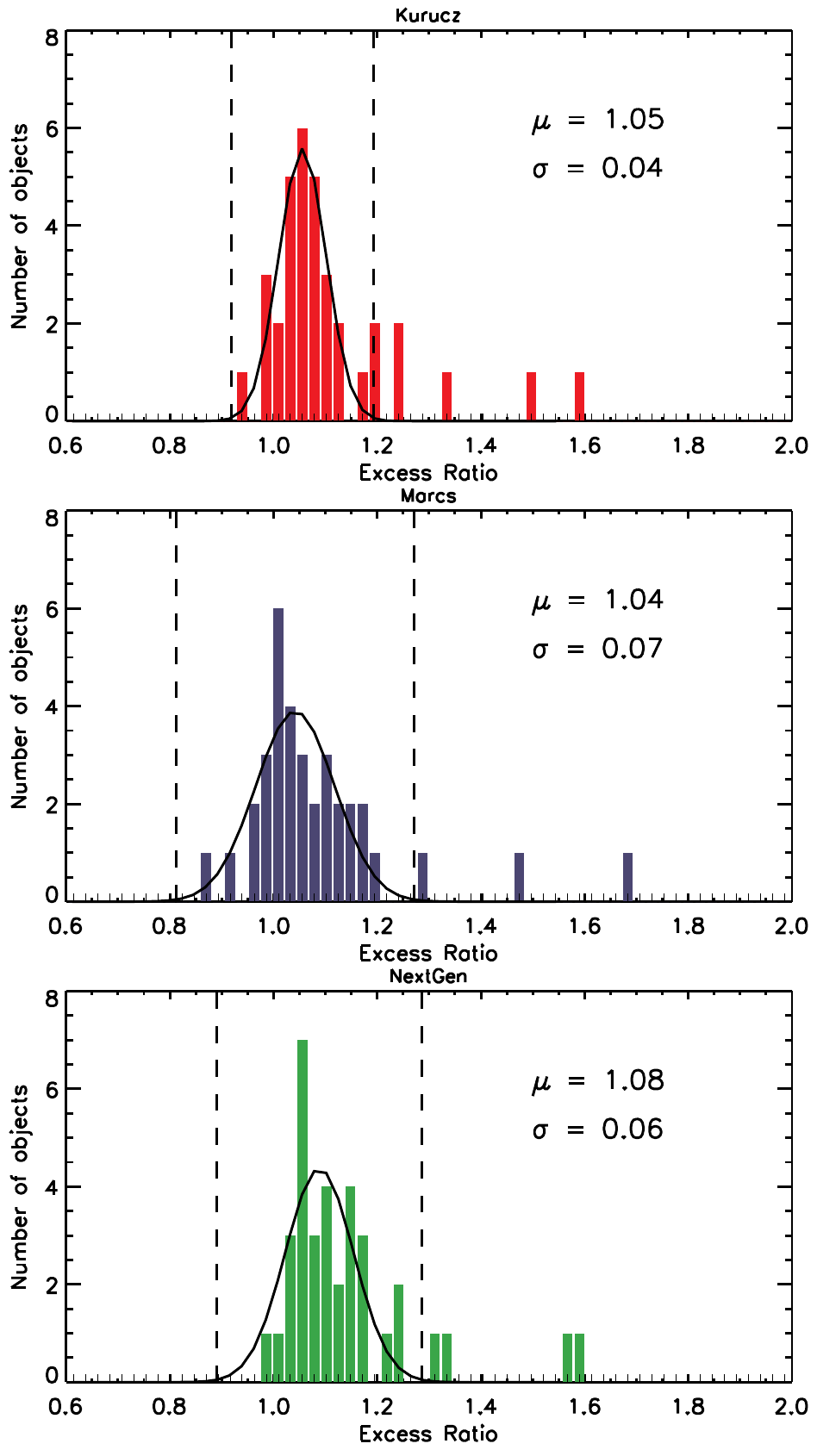}
    \caption{The distribution of excess ratios ($F_{obs}/F_{pred}$) at Spitzer MIPS $24$ $\mu$m for the
    sample of 37 stars using {\sc kurucz} (red), {\sc marcs} (blue)
    and {\sc NextGen} (green) model spectra as a means of predicting the
    stellar contribution.  Gaussian fits are shown as solid black
    lines and the x-translation and deviation parameters
    of the fit ($\mu$ and $\sigma$) are also displayed.  The $3\sigma$
    confidence intervals associated with the Gaussians are shown as
    the vertical dashed lines.  AKII437 and HII1132 (with very
    significant excess ratios) are omitted for the purpose of sensible
    axes scales. }
    \label{fig:hist}
\end{figure}

\noindent For each object, the excess ratio ($F_{obs}/F_{pred}$) was
calculated forming three distributions of excess ratios for each model
family.  These distributions were binned and the resulting histograms
are displayed in Fig. \ref{fig:hist}.  As shown, the excess ratios
form a gaussian-like distribution with positive outliers being candidate excesses.  It is immediately apparent that
these distributions are somewhat different despite representing the
same sample of stars.  Gaussian functions were fit to these
distributions in order to gain a quantitative measure of their width
and translation.  A narrow distribution with a translation of $\mu =
1.0$ describes model spectra which accurately predict the flux of
the target stars.  {While the {\sc marcs} distribution is best-centred on unity,
it is the broadest of the three distributions.  {\sc kurucz} model spectra form the narrowest distribution while the {\sc NextGen}
distribution has the largest offset from unity}.  These offsets have been previously noted of Pleiades stars - see Fig. 4 of \cite{gorlova_spitzer_2006}.

\noindent Detection of an infrared excess is considered conclusive
when the excess ratio of an object exceeds the upper $3\sigma$
confidence interval of the entire distribution (\cite{sierchio10}).
Excluding AKII 437 and HII 132 (which significantly outlie), it is
found that 7 objects (including 2 borderline objects) in the {\sc kurucz} distribution satisfy this
condition while falling to only 3 and 4 objects of the {\sc marcs} and
{\sc NextGen} distributions respectively.  There is agreement in all
model families that AKIA76 ($T_{\text{eff}} = 6310$ K, determined from the B-V colour), HII 1766 ($6720$ K) and PELS 146 ($5600$ K) have disk excesses.
AKII 383 ($6200$ K), PELS 20 ($5600$ K) and PELS 150 ($6200$ K)
yield disk excesses in using {\sc kurucz} model spectra while no
excesses exist if {\sc marcs} or {\sc NextGen} spectra are used.  In particular, the {\sc kurucz},
{\sc marcs} and {\sc NextGen} spectra whose synthetic photometry best
fit the photometry of AKII 383 have identical stellar parameters (T =
6000 K, log(g) = 4.5 , M/H = [0.0]).
\vspace{-0.75cm}

\section{Summary}

Detection of a disk can be concluded when infrared flux is in
excess of expectation from the star alone.  Popularly, a model spectra
is used to predict the stellar contribution and almost all previous
studies in this field have used {\sc kurucz} ({\sc ATLAS9}) model
spectra for this purpose.  {\sc marcs} and {\sc NextGen} model spectra
however have remained largely unused due to only recent availability.

\noindent An initial comparison of solar metallicity {\sc kurucz},
{\sc marcs} and {\sc NextGen} model spectra of identical stellar
parameters yielded differences as large as two orders of magnitude.
In extrapolating these spectra out to longer wavelengths, prediction
of Spitzer MIPS fluxes could differ by 5 \% for F, G and K spectra
types and as much as 15 \% for early M dwarfs.

\noindent A recent publication (\cite{sierchio10}) studied the Spitzer MIPS $24$
$\mu$m observations of 70 stars in the Pleiades cluster for possible
disk excesses, using {\sc kurucz} ({\sc ATLAS9}) models for prediction
of the stellar contribution.  This analysis was initially reproduced
for a subset of these stars and 7 out of 37 were
determined to have disk excesses.  In repeating this analysis using
{\sc marcs} and {\sc NextGen} model spectra, only 3 and 4
disk excesses were detected respectively.   {The differences between the model groups are such that different disk detection rates are obtained.  We encourage the use of more than one model family to determine the presence (or absence) of an infrared excess.} 
\vspace{-0.75cm}
%
%
%

\bibliographystyle{mn2e}
\bibliography{james}

%

\label{lastpage}

\end{document}